# Testing MOND with VIRGOHI21

Scott Funkhouser
*Department of Physics, Occidental College, 1600 Campus Rd., Los Angeles, CA 90047*
*Current address: W.M. Keck Science Center, 925N Mills Ave., Claremont, CA 91711-5916*

**ABSTRACT**
The "dark galaxy" VIRGOHI21 seems to be composed of an unusually high proportion of dark matter and is situated in a strong external gravitational field. As such it offers a rare test for theories of modified dynamics. If the system is bound then its dynamics are inconsistent with those predicted by the MOND theory.

VIRGOHI21 may be the first true "dark galaxy" discovered. The body appears to consist of $M_{HI} \sim 10^8$ solar masses of non-luminous hydrogen within a characteristic radius $R_{HI} \approx 16 \text{kpc} \approx 5 \times 10^{20}$m. The system exhibits a velocity width $\Delta V_{HI} \approx 220$km/s [1]. If these values are good characterizations of the system and if it is bound then its characteristic centripetal acceleration $a_{HI}$ would be

$$a_{HI} \sim \frac{\Delta V_{HI}^2}{R_{HI}} \sim 10^{-11} \text{ms}^{-2}. \qquad (1)$$

This implies that the dark galaxy features a dynamical mass of order $\Delta V_{HI}^2 R_{HI}/G \sim 10^{11}$ solar masses, where $G$ is the constant of gravitation [1]. VIRGOHI21 is also subject to the especially strong gravitational field of the Virgo cluster. These circumstances make the dark galaxy a rare test of theories such as MOND [2] that are meant to explain the observed dynamics of galaxies without relying on the presence of dark matter.

VIRGOHI21 is at least 120kpc from its closest neighboring galaxy [1] and is located about 820kpc from the center of the Virgo cluster. The characteristic peculiar velocities of the bound members of the cluster are near 1500km/s, a signature of the tremendous gravitational pull of Virgo. The gravitational field of the nearest neighboring galaxies at the location of VIRGOHI21 should be of order between $10^{-11}$ms$^{-2}$ and $10^{-12}$ms$^{-2}$. However, the total external gravitational field experienced by VIRGOHI21 is dominated by the field of the Virgo cluster, which is nearly of order $10^{-10}$ms$^{-2}$ at location of the dark galaxy. The external field may be approximated as being constant over the relatively small span of VIRGOHI21 and is roughly three orders of magnitude greater than the characteristic gravitational field $g_i$ generated by the body's baryonic mass given by

$$g_i \sim \frac{GM_{HI}}{R_{HI}^2} \sim 10^{-13} \text{ms}^{-2}. \qquad (2)$$

For a satellite system situated in a constant external gravitational field of magnitude $g_e$ that is greater than the magnitude of the internal gravitational field $g_i$ of the satellite, the MOND theory predicts that the mass elements of the satellite should exhibit quasi-Newtonian dynamics with an effective coefficient of gravitation $G_e$ given by

$$G_e \approx \frac{G}{\mu(g_e/a_0)} \qquad (3)$$

where $\mu(x)$ is the MOND interpolation function and $a_0$ is the MOND critical acceleration of order $10^{-10}$ms$^{-2}$ [3]. The function $\mu(x)$ rapidly approaches 1 for $x>1$ and rapidly approaches $x$ for $x<1$. In the case of VIRGOHI21 the magnitude of the external field is very close to the critical acceleration $a_0$, and MOND therefore prescribes that the body should exhibit dynamics that are quasi-Newtonian with $G_e \sim G$ without introducing invisible matter. That prescription is inconsistent with the inferred dynamics of the dark galaxy since the characteristic centripetal acceleration $a_{HI}$ is roughly two orders of magnitude greater than its characteristic baryonic gravitational field $g_i$. Presuming that VIRGOHI21 is rotationally supported, its dynamics are not explained by MOND alone. The dwarf spheroidal galaxies in Ursa Minor and Draco have been shown to feature similar disagreements with the MOND theory in that the modified dynamics do not alleviate the need for invisible mass [4].

ACKNOWLEDGEMENTS

This work benefited from many helpful discussions with M. Goldstein, S. McGaugh, M. Milgrom, and R. Minchin.

REFERENCES

[1] R. Minchin *et al.*, 2005, ApJ, 622, L21
[2] Milgrom M., 1983, ApJ, 270, 371
[3] Brada R., Milgrom M., 2000, ApJ, 541, 556
[4] Gerhard O., Spergel D., 1992, ApJ, 397, 38